\documentclass[12pt]{article}
\usepackage{graphicx,amsmath,amssymb}

\usepackage{epsf}
\usepackage{rotating}
\usepackage{epsfig}
\usepackage{supertabular}
\usepackage{subfigure}
\usepackage{multirow}

\def\rhoz{\rho^0}
\def\rhop{\rho^+}
\def\rhom{\rho^-}

\newcommand{\beq}{\begin{equation}}
\newcommand{\eeq}{\end{equation}}

\def\bmp#1{\begin{minipage}{#1\textwidth}}
\def\emp{\end{minipage}}

\newcommand\pubdate{\today}

\def\Cincy{Department of Physics, University of Cincinnati, Cincinnati, Ohio 45221,USA}

\def\Title#1{\begin{center} {\Large #1 } \end{center}}
\def\Author#1{\begin{center}{ \sc #1} \end{center}}
\def\Address#1{\begin{center}{ \it #1} \end{center}}

\newcommand\pubblock{\rightline{\begin{tabular}{l}
         \pubdate  \end{tabular}}}
\newenvironment{Abstract}{\begin{quotation}  }{\end{quotation}}
\newenvironment{Presented}{\begin{quotation} \begin{center} 
             PRESENTED AT\end{center}\bigskip 
      \begin{center}\begin{large}}{\end{large}\end{center} \end{quotation}}
\def\Acknowledgements{\bigskip  \bigskip \begin{center} \begin{large}
             \bf ACKNOWLEDGEMENTS \end{large}\end{center}}

\def\beq{\begin{equation}}
\def\eeq#1{\label{#1}\end{equation}}
\def\eeqn{\end{equation}}

\def\beqa{\begin{eqnarray}}
\def\eeqa#1{\label{#1}\end{eqnarray}}
\def\eeqan{\end{eqnarray}}

\let\bar=\overbar

\def\Dslash{\not{\hbox{\kern-4pt $D$}}}
\def\dslash{\not{\hbox{\kern-2pt $\del$}}}

\def\BR{\mbox{\rm BR}}

\def\msb{{\bar{\ssstyle M \kern -1pt S}}}

\input{babarsym}

\begin{document}
\begin{titlepage}
\pubblock

\vfill
\Title{\bf Unitarity triangle angles from penguin-dominated $B$ meson decays }
\vfill
\Author{Mathew T. Graham}
\Address{
SLAC National Accelerator Laboratory, Stanford, California 94309 USA}
\Author{Diego Tonelli}
\Address{ 
Fermilab, Batavia, Illinois 60510, USA
}
\Author{Jure Zupan\footnote{On leave of absence from FMF U. of Ljubljana, Jadranska 19, Ljubljana, Slovenia; IJS, Jamova 39, Ljubljana, Slovenia.}}
\Address{
\Cincy}
\vfill
\begin{Abstract}
In this time of transition to a new generation of quark  flavor experiments we review both the theoretical and the experimental progress on the determination of unitarity triangle angles from penguin-dominated $B$ decays. This summarizes the activities of the Working Group VI at the CKM2010 workshop.
\end{Abstract}
\vfill

\begin{Presented}
The 6th International Workshop on the CKM Unitarity Triangle, University of Warwick, UK, 6-10 September 2010
\end{Presented}
\vfill
\end{titlepage}
\def\thefootnote{\fnsymbol{footnote}}
\setcounter{footnote}{0}

\section{Introduction}
We are in a transition from the second to the third generation of flavor experiments. Even if still producing results, Belle and \babar\ have completed data taking and are being replaced by Belle-2 and SuperB, the next generation,  higher-precision B-Factories. CDF and D0, currently at the peak of their reach, will end operations in 2011, and leave the exploration of quark flavor dynamics in hadron collision to LHCb and (to a lower extent) ATLAS and CMS. 

A lasting legacy of this era is the experimental confirmation of the Kobayashi-Maskawa mechanism for CP violation in the Standard Model (SM), which is shown to describe data within $\approx 20\%$ uncertainties.

Intriguingly, some suggestive experimental hints that may not fit so well the CKM picture have emerged recently: the measurement of the branching fraction of $B^+\to \tau^+ \nu$ ~\cite{Asner:2010qj}, indications of a nonzero $B^0_s$ mixing phase ~\cite{Asner:2010qj},  and the anomalous like-sign dimuon asymmetry observed by D0~\cite{Abazov:2010hv} 
(see e.g. A. Soni \cite{SoniTalk}, and Refs. \cite{Ligeti:2010ia,Lenz:2010gu}). Because they involve observables precisely predicted by theory, further progress can mostly come from increased precision in experimental measurements. Furthermore, a plethora of other observables that can be equally important in either constraining New Physics (NP) models or in making discoveries is yet to be fully explored. Among these is the phenomenology of two-body $B$ decays,  which can test for the presence of NP in decay amplitudes as well as in flavor mixing. We review the theoretical and experimental state-of-the-art in this area, and the expected progress.

\section{Status of the theoretical predictions}
In the SM there is only one weak phase. Any two-body decay amplitude can thus be written as
\begin{equation}\label{decomp}
A(\bar B\to M_1 M_2)=e^{-i \gamma} T_{M_1 M_2}+P_{M_1M_2},
\end{equation}
where $\gamma$ is the same weak phase in all decays.
The $B$ decay amplitude is obtained by flipping the sign of the weak phase, $\gamma\to -\gamma$. It is customary to label the amplitude multiplying the weak phase as ``tree" ($T$), and the remaining part of the decay amplitude as ``penguin" ($P$). Both are complex in general, and carry strong phases. The relative sizes of tree and penguin amplitudes are governed by the CKM elements (absorbed in the definition of $T$ and $P$ above), the sizes of the relevant weak Hamiltonian Wilson coefficients,  and by the QCD dynamics. 

Reliable predictions of $T$ and $P$ are possible for decays into light mesons $M_{1,2}$,  in which case the hadronic elements factorize at leading order in the expansion in powers of the inverse $b$ quark mass  ($1/m_b$) \cite{Beneke:1999br,Bauer:2004tj}. The expression for the $B\to M_1 M_2$ amplitudes is schematically
\begin{equation}\label{schem}
A \propto 
T_{\zeta}\otimes \phi_{M_1} \otimes \zeta^{BM_2}+ T_{J} \otimes \phi_{M_1} \otimes \zeta_J^{BM_2} 
  +\dots,
\end{equation}
with $\otimes$ denoting the convolutions over momenta fractions, $\phi_{M}$ the light-cone distribution amplitude of a meson that does not absorb the spectator quark, $\zeta^{BM_2}$ the soft overlap function,   $\zeta_J^{BM_2}\propto \phi_B\otimes J\otimes \phi_{M_2}$ the function describing the completely factorizable contribution (including the jet function $J$), and $T_{\zeta,J}$ the corresponding hard scattering kernels,
while the ellipsis denotes higher orders and the potentially nonperturbative charming penguin contributions.

There are three theoretical schools of organizing the $1/m_b$ expansion: ``QCD factorization" (QCDF) \cite{Beneke:1999br}, ``Soft Collinear Effective Theory" (SCET) \cite{Bauer:2004tj} and ``perturbative QCD" (pQCD) \cite{Keum:2000wi}. All use the factorization at leading order in $1/m_b$, but make different choices for the treatment of expansion in $\alpha_S$, the treatment of $1/m_b$ corrections, the loops with charm quarks, and the choices for hadronic inputs (see a summary by G. Bell~\cite{BellTalk}).   For instance, there are three scales in the problem at or below $m_b$ mass: the hard scale, $m_b$, the hard-collinear scale, $\sqrt{m_b \Lambda_{\rm QCD}}$, and the nonperturbative scale, $\Lambda_{\rm QCD}$. The strong coupling $\alpha_S(m_b)$ at the hard scale is perturbative, while $\alpha_S(\Lambda_{\rm QCD})$ is nonperturbative. QCDF and pQCD expand in $\alpha_S(\sqrt{\Lambda_{\rm QCD} m_b})$, while SCET does not, introducing rather new nonperturbative parameters. This choice becomes unpractical beyond NLO in $\alpha_S(m_b)$ because too many parameters will be introduced. Luckily the expansion in $\alpha_S(\sqrt{\Lambda_{\rm QCD }m_b})$ seems to be well behaved so that the introduction of new unknown parameters from this source may be avoided. 

The perturbative calculations have been done at NLO in SCET \cite{Jain:2007dy}, and only partially at NLO in pQCD. In QCDF two groups are performing the perturbative calculations at NNLO, with several results already available \cite{Bell:2007tv}. Because of a potentially large missing piece in a formally  $1/m_b$ suppressed- but chirally-enhanced term, an effort to complete the NNLO calculation is ongoing (in QCDF counting which treats 1-loop corrections to spectator scattering as NNLO). The NNLO calculations have been completed for the tree amplitudes and the perturbative expansion seems well behaved. In particular, factorization was found to explicitly hold at this order, as expected. Because of the cancellations between vertex and spectator scattering, the NNLO predictions for branching ratios are very similar to the NLO ones. One now has precise predictions on color-allowed tree amplitudes that are in agreement with data. The color-suppressed tree amplitudes, on the other hand, suffers from hadronic uncertainties. An important parameter here is the first inverse moment of the $B$ meson light-cone distribution amplitude, $\lambda_B^{-1}$. The data seem to prefer a smaller value of $\lambda_B$. As stressed by G. Bell \cite{BellTalk} further refinements are possible with improved experimental inputs: $\lambda_B$ can be determined from $B^+\to l^+\nu \gamma$ with an energetic photon, the $B^+\to \rho l^+ \nu$ spectrum can be used to determine $|V_{ub}| A_0^{B\rho}(0)$, while very useful information can come from tree-dominated $B^0_s$ decays and especially pure annihilation decays $B^0\to K^+K^-$, $B^0_s\to \pi^+\pi^-, \pi^+\rho^-, \rho^+\rho^-$. 

A key question on which the three schools disagree is on the source of the fairly large values of strong phases observed. In QCDF the phase comes from $1/m_b$ corrections, with a fit to data made using a crude phenomenological model for the nonperturbative weak annihilation. SCET on the other hand postulates that charming penguins are nonperturbative and carry a strong phase.  Following Ref.~\cite{Beneke:2009az} some view the issue fully resolved in favor of a perturbative charming penguin \cite{JagerTalk}. Phenomenologically, the two possibilities will be very hard to distinguish since they contribute in the same way to the decay amplitudes. This issue may still stay with us for some time. 

The pQCD school uses $k_T$ factorization to factorize even the soft overlap function that receives significant contributions from soft or small-$x$ physics. The claim is that the strong phase is coming from annihilation as in QCDF, and is also perturbatively calculable. Another source of strong phases was found recently from uncanceled Glauber divergences in spectator-scattering amplitudes leading to a universal nonperturbative phase~\cite{MishimaTalk}.
~The effect is potentially large for color-suppressed tree amplitudes,  yielding a large strong phase, while it is small for color-allowed tree and penguin amplitudes. A potential concern, since this is a universal phase, is that it will spoil the agreement of data with predictions of $B$ meson decays into pseudoscalar-vector or vector-vector final states ($B\to PV, VV$).

A real roadblock to the precision application of the $1/m_b$ expansion seem to be the poorly understood $1/m_b$ terms. By going beyond the crude models employed by QCDF one will quite likely lose predictivity,  given the plethora of new and unknown hadronic matrix elements that arise at subleading order in $1/m_b$. Furthermore, at subleading powers in $1/m_b$, the short--long distance factorization breaks down. As warned by A. Kagan,  the amplitudes could be dominated by soft and nonperturbative physics~\cite{KaganTalk}.~
The first indications could be already the infrared logarithmic divergences seen in the convolution integrals. If taken seriously, the dominant meson production occurs in an asymmetric configuration with one fast and one soft valence constituent. Large soft overlaps are being seen in CLEO-c continuum $e^+e^-\to \pi^+\pi^-, K^+K^-$ data, where the soft overlap power corrections to the form factors are an order of magnitude larger than the perturbative part. Improving the $O(\Lambda_{\rm QCD}/m_b)\sim O(20\%)$ predictions that we have now, thus, seems extremely challenging, if not impossible. 

There are ways, however, to use the $1/m_b$ expansion, even if the accuracy of predictions is limited to $20\%$. The general strategy is to identify observables that are precisely predicted by independent constraints, for instance by using flavor symmetry arguments. The expansion in $1/m_b$ could then be used to calculate the corrections to these predictions. An example are correlations between CP-violating asymmetries $S_{K_S\pi^0}$ and $C_{K_S\pi^0}$ obtained in Ref.~\cite{Fleischer:2008wb}, where QCDF was used only to estimate the size of the SU(3) breaking terms. There are other examples of employing flavor symmetries to extract interesting weak scale physics information from two body $B$ decays that could benefit from QCDF/SCET predictions to constrain the uncertainties only. A well known one is the use of flavor SU(3) to obtain information on $\gamma$ 
from $B^0\to \pi^+\pi^-$ and $B^0_s\to K^+K^-$ decays, where Fleischer and Knegjens obtain $\gamma=(68.5^{+4.5}_{-5.8}{}^{+5.0}_{-3.7}{}^{+0.1}_{-0.2})^\circ$ with uncertainties due to the experimental inputs, the current estimate of the magnitude of SU(3) breaking (not using QCDF), and the estimates of the SU(3) breaking phase~\cite{FleischerTalk}. Another useful strategy is the measurement of the standard unitarity triangle angle $\beta$ using penguin dominated modes. We next review the experimental status of these efforts.

\section{Measurements of $\beta$ from charmless decays}
\label{Beta}

The study of $\beta$ in charmless $B$ decays focuses on comparing $\beta_{\mathit{SM}}$ measured in the ``golden'' charmonium modes (e.g. J/$\psi\KS$) with $\beta_{\mathit{eff}}$ measured in penguin-diagram dominated decays (e.g. $\phi\KS$).   To first order, the time-dependent CP violating (CPV) parameters $S$ and $C$ for charmless penguin decays should be the same as in the golden modes, namely $-\eta_f S_f \approx \sin{2\beta_{\mathit{SM}}}$ and $C\approx 0$.  There are small final-state--dependent corrections due to CKM-suppressed SM tree contributions (cf. Eq. \eqref{decomp}).
More interestingly, heavy particles from beyond-SM sources (e.g. supersymmetric sector or a fourth quark generation) may participate in the loops and shift the observed value of the CPV parameters.  

The B-Factories have performed extensive studies of a large number of penguin-dominated charmless $B$ decays.  For modes such as $\Bz\to\eta^\prime\KS$ or $\Bz\to\piz\KS$, the final state is treated in a ``quasi-two body'' (Q2B) way and the experimental analysis proceeds in the same way as in $\Bz\to J/\psi\KS$ \cite{:2009yr,Chen:2006nk}.   Three-body final states, however, are analyzed using a time-dependent Dalitz plot technique,
which resolves the trigonometric ambiguity inherent in the Q2B method.  The Dalitz plot technique also accounts for interference effects between different resonances which otherwise can be a significant source of systematic uncertainty in the CPV parameters.  

\begin{figure}[tb]
\includegraphics[width=0.400\textwidth,angle=0]{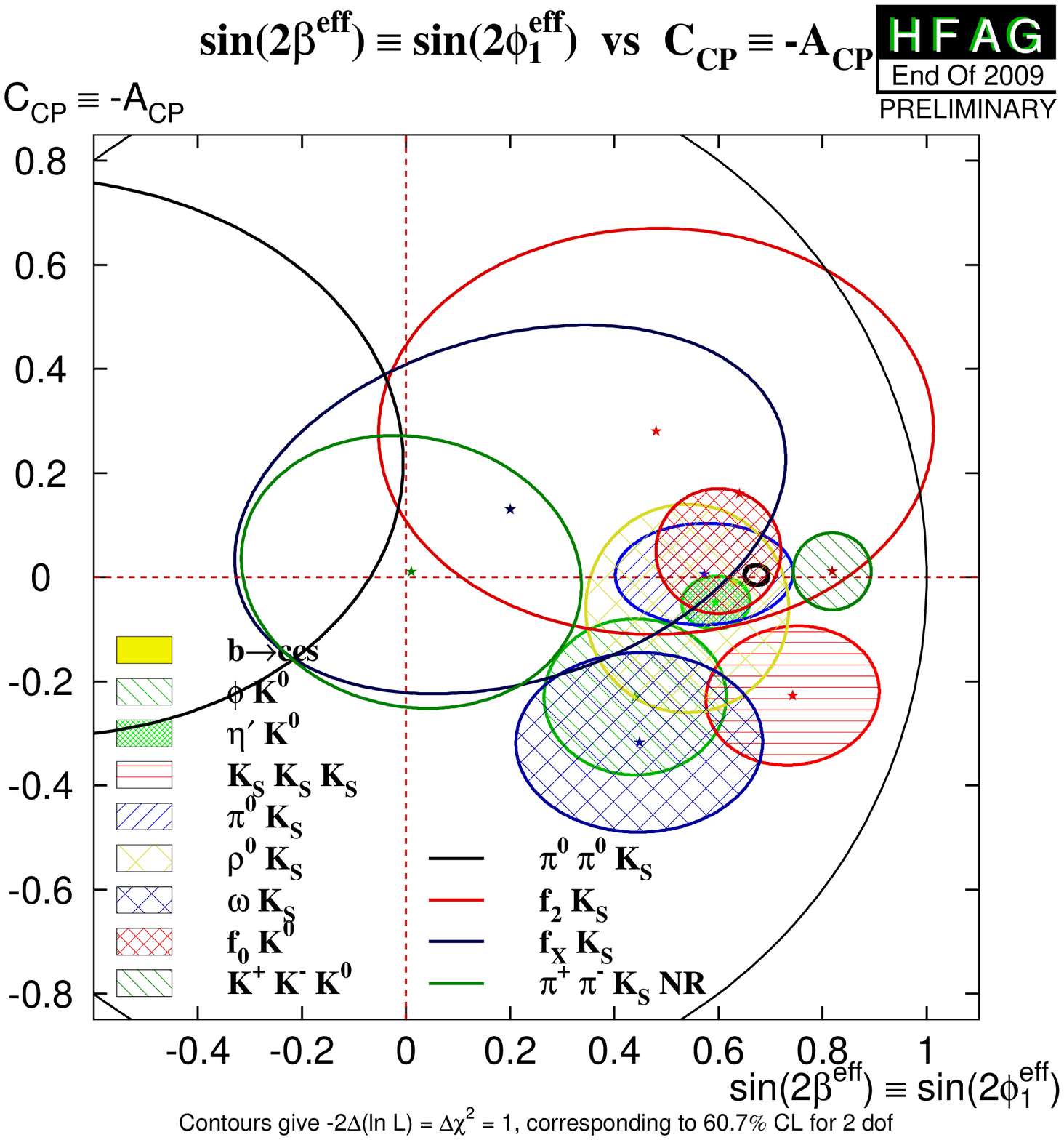}
\includegraphics[width=0.600\textwidth,angle=0]{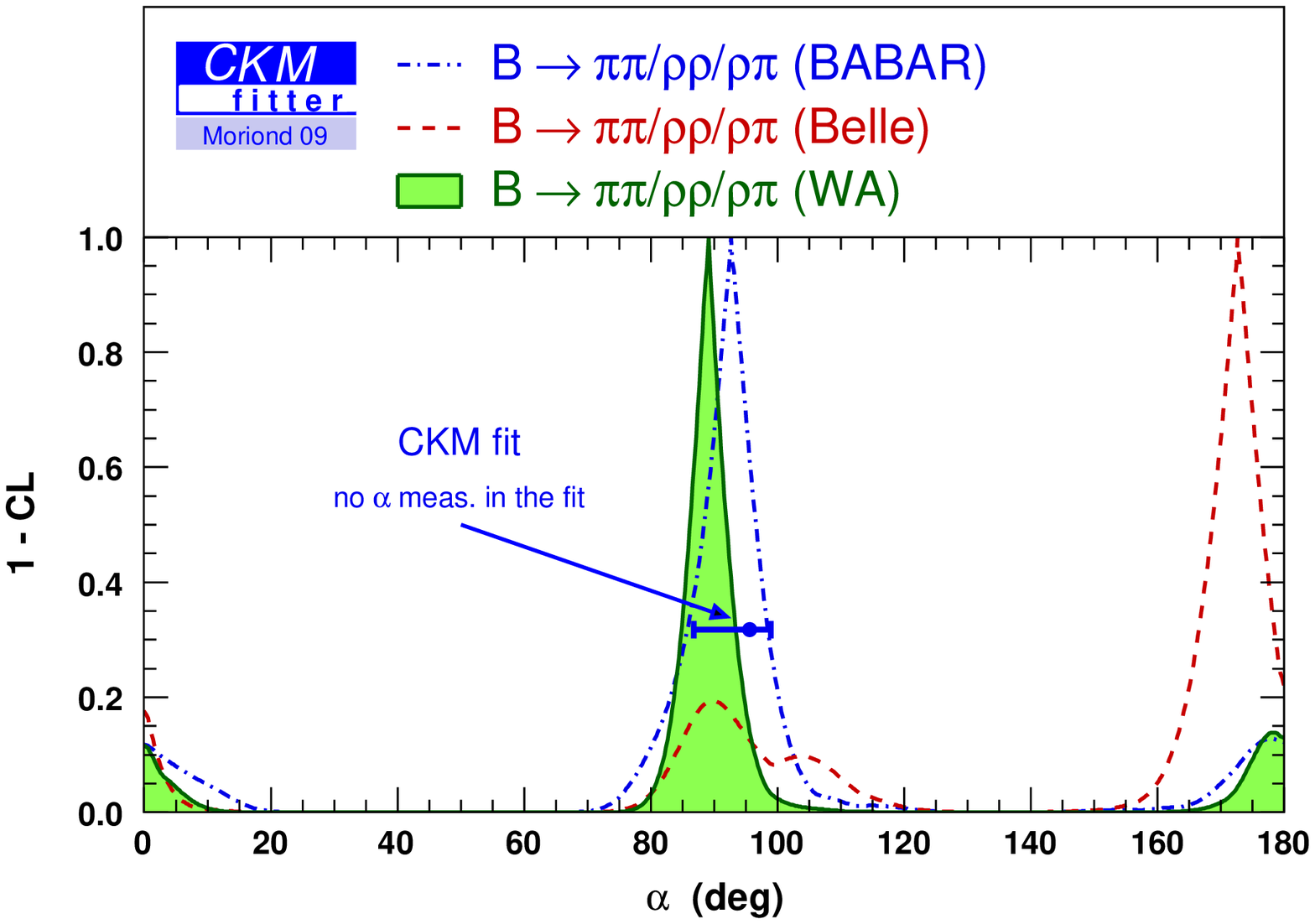}
\caption{\label{fig:SvsC}
Measured  $S$ as a function of $C$ for  penguin-dominated $B$ decays (left). The values measured from charmonium decays are shown in the shaded yellow.
Profile of 1-CL as a function of the angle $\alpha$ from $\pi\pi$, $\rho\pi$, and $\rho\rho$ decays (right).}
\end{figure}

A summary of the current results for $S=\sin{2\beta_{\mathit{eff}}}$ (or $\beta_{\mathit{eff}}$) and $C$ from penguin-dominated decays is shown in Table \ref{tab:resultsSumm} and Figure \ref{fig:SvsC} (left).  Individually, none of the measurements deviate from the SM by more than two standard deviations and the naive average of all $\sin{2\beta_{\mathit{eff}}}$ measurements  
(valid in the limit of vanishing tree contributions in all the modes) agrees with the charmonium-based determinations within one standard deviation.  The uncertainties  on the measurements are dominated by their statistical component  and the majority of the systematic uncertainties should scale with sample size, which bodes well for the future reach at the next generation of B-Factories.  With 75$\invab$ of data, many of these observables are expected to be measured to a few percent \cite{Manoni:2011mj}.  

\begin{table}[thb]
\centering
\begin{tabular}{llcc}
\hline
\hline  
  Decay      &   Ref.      & $-\eta_{f}S_{f}=\sin{2\beta_{\mathit{eff}}}$  & $C_{f}=-A_{f}$ [\%]    \\

\hline
$\Bz\to c\overline{c} K^0$    & HFAG~\cite{Asner:2010qj}  &  $0.672\pm0.023$   &   $0.4 \pm 1.9  $           \\
  $\Bz\to\eta^{\prime}K^0$    & \babar~\cite{Aushev:2011cc,:2008se} &  $0.57\pm0.08\pm0.02$          &       $-8\pm6\pm2$    \\
                       & Belle~\cite{Aushev:2011cc,Chen:2006nk} &  $0.64\pm0.10\pm0.04$          &       $1\pm7\pm5$    \\
  $\Bz\to\omega \KS     $    & \babar~\cite{Aushev:2011cc,:2008se} &  $0.55^{+0.26}_{-0.29}\pm0.02$   &      $-52^{+22}_{-20}\pm3$      \\
                       & Belle~\cite{Aushev:2011cc,Abe:2006gy} &  $0.11\pm0.46\pm0.07$           &      $9\pm29\pm6$       \\
  $\Bz\to\pi^0 K^0     $    & \babar\footnote{\babar only uses the $\piz\KS$ final state.}~\cite{Aushev:2011cc,:2008se}
                                    &   $0.55\pm0.20\pm0.03$          &       $13\pm13\pm3$     \\
                       & Belle~\cite{Aushev:2011cc,Fujikawa:2008pk} &    $0.67\pm0.31\pm0.08$          &       $-14\pm13\pm6$   \\
\hline
             &                      & $ \beta_{\mathit{eff}}$  &   \\
\hline
&&\\[-0.2cm]
$\Bz\to c\overline{c} K^0$    & HFAG~\cite{Asner:2010qj}  &  $(21.1\pm0.09)^\circ$   &   $0.4 \pm 1.9  $            \\
  $ \Bz\to\phi K^0     $    & \babar~\cite{Miyabayashi:2011nx,:2008gv} & $(7.7\pm7.7\pm0.9)^\circ$   &   $14\pm19 \pm 2  $          \\
                       & Belle~\cite{Miyabayashi:2011nx,Nakahama:2010nj} &    $(32.2\pm9.0\pm3.0)^\circ$   &   $4 \pm 20\pm10  $           \\
  $ \Bz\to \KpKm K^0$ (no $\phi/f_0$)       
                       & \babar~\cite{:2008gv} &   $(29.3\pm4.6\pm1.7)^\circ$   &   $5\pm9 \pm 4  $          \\
  
                     & Belle~\cite{Abe:2006gy} &    $(24.9\pm6.4\pm3.3)^\circ$   &   $-14\pm11 \pm 9  $          \\
  $ \Bz\to f_0(980)(\to K^+K^-) K^0     $    
                       & \babar~\cite{Miyabayashi:2011nx,:2008gv} &  $(8.5\pm7.5\pm1.8)^\circ$   &   $1\pm26 \pm 7  $         \\
  & Belle~\cite{Miyabayashi:2011nx,Nakahama:2010nj} &  $(31.3\pm9.0\pm5.2)^\circ$   &   $-30\pm29 \pm 14  $     \\
  $ \Bz\to f_0(980)(\to\pi^+\pi^-) \KS     $    
                        & \babar~\cite{Miyabayashi:2011nx,Aubert:2009me}  &   $(36.0\pm9.8\pm3.0)^\circ$   &   $-8\pm19 \pm 5  $          \\
  & Belle~\cite{Miyabayashi:2011nx,:2008wwa} &    $(12.7^{+6.9}_{-6.5}\pm4.3)^\circ$   &   $-6\pm17 \pm 11  $       \\
  $ \Bz\to \rho \KS     $    
                       & \babar~\cite{Miyabayashi:2011nx,Aubert:2009me}  &   $(10.2\pm8.9\pm3.6)^\circ$   &   $5\pm26 \pm 10  $          \\
                       & Belle~\cite{Miyabayashi:2011nx,:2008wwa}  &    $(20.0^{+8.6}_{-8.5}\pm4.7)^\circ$   &   $3^{+23}_{-24} \pm 15  $          \\
\hline
\hline
\end{tabular}
    \caption{\label{tab:resultsSumm}
        Time-dependent CPV parameters measured in penguin dominated $B$ decays.  The three-body modes (lower table) provide sensitivity to both the sine and cosine terms, resolving the trigonometric ambiguity and allowing extraction of  $\beta_{\textit{eff}}$.   
}
  \end{table}



 \section{Measurements of $\alpha$ from charmless decays}
\label{Alpha}

Tree-level amplitudes of charmless $B$ decays such as $B^0\to\pi^+\pi^-$ are sensitive to the unitarity triangle angle $\alpha$.   Penguin contributions to the decays complicate the extraction of $\alpha$ and various techniques have been devised to account for them.  For decays into CP-eigenstates like $B\to\pi\pi$ and $B\to\rho\rho$, $\alpha$ can be extracted using an isospin decomposition \cite{Snyder:1993mx}.  This technique requires the measurement of all observables of all charge combinations.  For the three-body final state $B^0\to\rho\pi\to\pi^+\pi^-\pi^0$, a full time-dependent amplitude analysis is needed~\cite{Aubert:2007jn,Kusaka:2007dv}.

The measurements used for the isospin extraction of $\alpha$ from $\pi\pi$ and $\rho\rho$ final states are given in Table \ref{tab:pipirhorhoSumm}.  The $\pipi$ decays show evidence for large hadronic penguin amplitudes, manifested in the large $C$ term in  $\Bz\to\pip\pim $, and thus the sensitivity to $\alpha$ is weak; with the current measurements the 90\% exclusion region is only [11$^{\circ}$,79$^{\circ}$].   On the other hand, the  $\rho\rho$ final state, which is nearly 100\% longitudinally polarized, is dominated by the tree amplitude with very little penguin pollution. The $\alpha$ obtained from combining the $B\to\rho\rho$ observables is $(89.9\pm5.4)^{\circ}$  in the region favored by the global 
unitarity triangle fit. 

\begin{table}[thb]
\centering
\begin{tabular}{llccc}
\hline
\hline  
Decay			&	Ref.									&	$\mathcal{B}$ ($\times 10^{6}$)	&	$\eta_{f}S_{f}=\sin{2\alpha_{\mathit{eff}}}$	&	$C_{f}=-A_{f}$ [\%]	\\
\hline
$\Bz\to\pip\pim $	& 	\babar~\cite{Dalseno:2011uu,Aubert:2008sb}	& 	$5.5\pm0.4\pm0.3$  				&	$-0.68\pm0.10\pm0.03$   &      $-25\pm80\pm2$       \\
                       & Belle~\cite{Dalseno:2011uu,Ishino:2006if} &  $5.1\pm0.2\pm0.2$   &  $-0.61\pm0.10\pm0.04$           &      $-55\pm8\pm5$     \\
  $\Bp\to\pip\piz$    & \babar~\cite{Dalseno:2011uu,Aubert:2007hh} &   $5.0\pm0.5\pm0.3$   &  --          &       $-8\pm6\pm2$    \\
                       & Belle~\cite{Dalseno:2011uu,Abe:2006qx} &   $6.5\pm0.4\pm0.4$  &  --          &       $1\pm7\pm5$     \\
  $\Bz\to\piz\piz $    & \babar~\cite{Aubert:2008sb} &   $1.8\pm0.2\pm0.1$ &     --   &      $43\pm25\pm5$       \\
                       & Belle~\cite{Abe:2006cx} &   $1.1\pm0.3\pm0.1$  &      --       &      $44^{+73}_{-62}~^{+4}_{-6}$       \\
\hline
  $\Bz\to\rhop\rhom$   & \babar~\cite{Dalseno:2011uu,Aubert:2007nua} & $25.5\pm2.1^{+3.6}_{-3.9}$   &  $-0.17\pm0.20\pm0.06$    &  $-1\pm15\pm6$    \\
                       & Belle~\cite{Dalseno:2011uu,Abe:2007ez} & $22.8\pm3.8^{+2.3}_{-2.6}$    &  $0.19\pm0.30\pm0.07$     &  $-16\pm21\pm7$   \\
  $\Bp\to\rhop\rhoz$    &\babar~\cite{Dalseno:2011uu,Aubert:2009it} & $23.7\pm1.4\pm1.4$  &  --   &  $5\pm6\pm1$    \\
                       & Belle~\cite{Dalseno:2011uu,Zhang:2004wza} &  $31.7\pm7.1^{+3.8}_{-6.7}$   &   --          &       $1\pm7\pm5$    \\
  $\Bz\to\rhoz\rhoz$   & \babar~\cite{Dalseno:2011uu,:2008iha} &  $0.9\pm0.3\pm0.1$  &  $0.3\pm0.7\pm0.2$    &  $20\pm80\pm30$      \\
                       & Belle~\cite{Dalseno:2011uu,:2008et} &  $0.4\pm0.4^{+0.2}_{-0.3}$   &  $0.11\pm0.46\pm0.07$ &      $9\pm29\pm6$      \\
\hline
\hline
\end{tabular}
    \caption{\label{tab:pipirhorhoSumm}
        Summary of the $B\to\pi\pi$ and $B\to\rho\rho$ measurements used to extract $\alpha$.
}
\end{table}



The $\rho\rho$ result dominates the average, although $\pi\pi$ does contribute significantly and $\rho\pi$ excludes the solution at (180-$\alpha$)$^\circ$.   One point to understand is that the precision is highly dependent on the individual values of the measurements in the isospin decompositions and thus, even though the uncertainties of the measurements will decrease with additional luminosity, the error on $\alpha$ likely will not scale as nicely. The current $\rho\rho$ isospin triangle is  stretched because of a large value of $\mathcal{B}(B^+ \to \rho^+\rho^0)$  dominated by the precision of BaBar measurement.  An updated measurement from Belle could be extremely useful. At precision of a few degrees one also needs to  worry about isospin symmetry breaking effects. These are harder to quantify, but are likely to be smaller in $\alpha$ extracted from $B\to \rho \pi$ \cite{Gronau:2005pq}.

An additional method to extract $\alpha$, proposed by Gronau and Zupan~\cite{Gronau:2005kw}, suggests to use $B\to a_1\pi,~a_1 K,~K_1\pi$ and relate them using SU(3) symmetry (this method is applicable also to $B\to \rho\pi$ \cite{Gronau:2004tm}).  Since $a_1^\pm \pi^\mp $ is not a CP-eigenstate, the additional parameters $\Delta C$ and $\Delta S$ are needed to describe the time evolution.  The former describes the asymmetry between the sum of rates $(B^0 \to a_1^+\pi^- + \overline{B}^0\to a_1^-\pi^+)$ and 
$(B^0 \to a_1^-\pi^+ + \overline{B}^0\to a_1^+\pi^-)$,  while the latter is related to the strong phase difference between amplitudes; neither of these parameters are sensitive to CP-violation.  Between \babar\  and Belle, all of the necessary measurements have been performed (see Table \ref{tab:a1piSumm})~\cite{Stracka:2011ag}.~
 The extraction of the $B\to K_{1}\pi$ branching fraction is complicated by the fact that the SU(3) partner of the $a_1$ is a mixture of the $K_{1}(1270)$ and $K_{1}(1400)$ states.  Assuming SU(3), the value of $\alpha$ extracted from this method is $(79\pm7\pm11)^\circ$, where the first uncertainty is the combination of statistical and systematic contributions and the second is due to penguin pollution.  

\begin{table}[thb]
  \begin{center}
\begin{tabular}{lcc}
\hline
\hline  
Decay        &   Quantity      &      Value\\

\hline
&&\\[-0.2cm]
  $\Bz\to a_1^{\pm}\pi^{\mp} $    &          $\mathcal{B}$ (\babar)        &         $(33.2 \pm 3.8 \pm 3.0) \times 10^{-6}$   \\
                                 &          $\mathcal{B}$ (Belle)         &         $(29.8 \pm 3.2 \pm 4.6) \times 10^{-6}$   \\
                                 &          $A_{CP}$         &   $-0.07\pm0.07\pm0.02$          \\
                                 &          $S$         &     $0.37\pm0.21\pm0.07$         \\
                                &          $\Delta S$         &   $-0.14\pm0.21\pm0.06$           \\
                                &          $C$         &     $-0.10\pm0.15\pm0.09$         \\
                                &          $\Delta C$         &    $0.26\pm0.15\pm0.07$          \\
  $\Bz\to a_1^{-} \Kp $         &          $\mathcal{B}$        &         $(16.4 \pm 3.0 \pm 2.4) \times 10^{-6}$   \\
  $\Bp\to a_1^{+} K^{0} $         &         $\mathcal{B}$       &         $(34.8 \pm 5.0 \pm 4.4) \times 10^{-6}$   \\
  $\Bz\to K_{1}(1270)^+\pim + K_{1}(1400)^{+}\pim $         &          $\mathcal{B}$        &         $(31^{+8}_{-7}) \times 10^{-6}$   \\
  $\Bp\to K_{1}(1270)^0\pip + K_{1}(1400)^{0}\pip $         &          $\mathcal{B}$       &         $(29^{+29}_{-17}) \times 10^{-6}$   \\
\hline
\hline
\end{tabular}
    \caption{\label{tab:a1piSumm}
        Measurements used to extract $\alpha$ from the $a_1 \pi$ and SU(3)-related modes~\cite{Stracka:2011ag}.
}
    \vspace{-0.8cm}
  \end{center}
\end{table}

 \section{Measurements of $\gamma$ from charmless decays}

The determination of  $\gamma$ through charmless decays may supplement the still limited precision attained through the tree-dominated $B^{\pm}\to DK^{\pm}$  decays \cite{WGV} and offer a sensitive probe of non-SM contributions. The $V_{ub}$ phase enters the amplitudes of $B^0$ and $B^0_s$ decays into light hadrons (see Eq.~\eqref{decomp}). Large penguin pollution and potentially significant contributions of sub-leading topologies (e.g. annihilation) prevent a precise extraction of $\gamma$ from a single decay mode. However, these loop-induced processes may receive contributions from virtual exchange of heavy non-SM particles, which can modify the observed branching ratios or CP-violating asymmetries  from the SM-expected values. Theory predictions suffer from hadronic uncertainties and have generally worse precision than experimental measurements. Combinations of observables from multiple channels related by isospin or SU(3) flavor symmetry are useful to extract information on $\gamma$ and constrain non-SM physics.

Belle and \babar\ recently studied the interference pattern in Dalitz plots of  $B^0\to K^0_S \pi^+\pi^-$ and $B^0\to K^+ \pi^-\pi^0$ decays, where penguin contributions are suppressed or corrected by using symmetry arguments~\cite{PuccioTalk}. The results are somewhat inconsistent despite large uncertainties dominated by the statistical contribution. A clearer picture will be provided by more copious samples available at LHCb and super-flavor factories.
In the past decade CDF has been pioneering the exploration of two-body $B^0_s$ decays with first measurements of asymmetries and branching ratios of $B^0_s \to K^+K^-$ and $B^0_s \to K^-\pi^+$ decays~\cite{Abulencia:2006psa, Aaltonen:2008hg, Aaltonen:2011qt} which, combined with $B^0$ results,  allow extraction of $\gamma$~\cite{FleischerTalk}. No penguin annihilation mode, $B^0_s \to\pi^+\pi^-$  or $B^0\to K^+K^-$,  has yet been observed but upper bounds on their decay rates have been greatly improved. CDF has been joined recently by Belle~\cite{LeeTalk} and LHCb~\cite{CarboneTalk} but no analysis of time-evolution has been pursued thus far. Sample size is still a limiting factor once production flavor needs to be identified. See Table~\ref{tab:sec3} for a few recent results. First LHCb data are promising toward a time-dependent analysis in the near future. On a longer term,  LHCb plans also to study the combined time-independent Dalitz plots of $B^+\to K^+\pi^-\pi^+$ and $\B^0 \to K^0_S \pi^+\pi^-$ decays, which provide information on $\gamma$.
\begin{table}[thb]
\centering
\begin{tabular}{llcc}
\hline
\hline
Decay				&	Ref.		&	$\mathcal{B}$ ($\times 10^{6}$)		&	$A_{CP}$ [\%]			\\
\hline
$B^0 \to K^+\pi^-$		&	\babar~\cite{Aubert:2006fha,Aubert:2008sb} 	&	$19.1 \pm 0.6 \pm 0.6$			&	$-10.7 \pm 1.6 \pm 0.6$	\\
					&	Belle	~\cite{Abe:2006qx,:2008zza}			&	$19.9 \pm 0.4 \pm 0.8$			&	$-9.4 \pm 1.8 \pm 0.8$	\\
					&	CDF~\cite{Aaltonen:2011qt}				&	--							&	$-8.6 \pm 2.3 \pm 0.9$					\\
$B^+ \to K^+\pi^0$		&	\babar~\cite{Aubert:2007hh} 				&	$13.6 \pm 0.6 \pm 0.7$			&	$3.0 \pm 3.9 \pm 1.0	$	\\
					&	Belle~\cite{Abe:2006qx,:2008zza} 			&	$12.4 \pm 0.5 \pm 0.6$			&	$7.0 \pm 3.0 \pm 1.0$		\\
$B^+ \to K^0\pi^+$		&	\babar~\cite{Aubert:2006gm} 				&	$23.9 \pm 1.1 \pm 1.0$			&	$-2.9 \pm 3.9 \pm 1.0$	\\
					&	Belle~\cite{Abe:2006xs} 					&	$22.8 \pm 0.8 \pm 1.3$			&	$3.0 \pm 3.0 \pm 1.0	$	\\	
$B^0 \to K^0\pi^0$		&	\babar~\cite{Aubert:2008sb,:2008se} 		&	$10.1 \pm 0.6 \pm 0.4$			&	$-13 \pm 13 \pm 3$		\\
					&	Belle~\cite{Fujikawa:2008pk} 				&	$8.7 \pm 0.5 \pm 0.6$			&	$14 \pm 13 \pm 6$		\\
$B^0_s\to K^+K^-$		&	CDF~\cite{Aaltonen:2011qt} 			&	$23.9 \pm 1.4 \pm 3.6$			&	--				\\
					&	Belle	~\cite{Peng:2010ze}					&	$38 \pm 10\pm 7$				&	--				\\
$B^0_s\to K^-\pi^+$		&	CDF~\cite{Aaltonen:2008hg} 			&	$5 \pm 0.7 \pm 0.8$				&	$39 \pm 15 \pm 8$	\\
					&	Belle~\cite{Peng:2010ze} 				&	$<$ 26 at the 90\% CL			&					\\
$B^0_s\to \pi^-\pi^+$	&	CDF~\cite{Aaltonen:2008hg}				&	$<$1.2 at the 90\% CL			&	--				\\
					&	Belle	~\cite{Peng:2010ze}					&	$<12$ at the 90\% CL			&	--				\\
$B^0_s\to K^0 \bar{K}^0$	&	Belle	~\cite{Peng:2010ze}					&	$<66$ at the 90\% CL			&	--				\\		
$B^0 \to K^{*0} \bar{K}^{*0}$	&	Belle	~\cite{Chiang:2010ga}				&	$<0.8$ at the 90\% CL			&	--				\\			
$B^0 \to K^{*0}  K^{*0}$		&	Belle	~\cite{Chiang:2010ga}				&	$<0.2$ at the 90\% CL			&	--				\\	
$B^+\to \eta ' \rho^+$	&	\babar~\cite{delAmoSanchez:2010qa}		&	$9.7\pm 1.9 \pm 1.1$			&	--				\\
$B^+\to a_1^+K^{*}(892)^0$	& \babar~\cite{Sanchez:2010qm}			&	$<3.6$ at the 90\% CL			&	--				\\
$B^+	\to K^+\pi^0\pi^0$	&	\babar~\cite{delAmoSanchez:2010fz}		&	$15.5 \pm 1.1 \pm 1.6$			&	--				\\
$B^0\to \pi^+K_s^0K^-$	&	\babar~\cite{delAmoSanchez:2010ur}		&	$3.2 \pm 0.5 \pm 0.3$			&	--				\\
\hline
\hline
\end{tabular}
    \caption{\label{tab:sec3}
        Branching fractions and CP-violating asymmetries of decays sensitive to $\gamma$ or useful to test the $1/m_b$ expansion.}
\end{table}

 \section{Tests of the 1/$m_b$ expansion}
The pattern of direct CP violation in charmless $B$ meson decays shows some unanticipated discrepancies from SM predictions that use $1/m_b$ expansion (Table~\ref{tab:sec3}). If strong phases carried by tree amplitudes are small (or if the color suppressed tree amplitude is small), then CP asymmetries for $B^0\to K^+\pi^-$ and $\Bu\to K^+\pi^0$ decays should be similar~\cite{Beneke:1999br,Keum:2002vi,Gronau:1998ep,Gronau:2006ha}. However, experimental data show a significant discrepancy~\cite{:2008zza,Aubert:2007hh},  which has prompted intense activity because several simple SM extensions could naturally accommodate the discrepancy. The emerging picture suggests that the effect is likely due to the presence of large CP-conserving phases \cite{Gronau:2008gu}. These are predicted to be small in QCDF and SCET uses of $1/m_b$ expansions, while a potential source of large strong phase is claimed be identified in pQCD~\cite{MishimaTalk}. The anomaly persists irrespective of $1/m_b$ corrections only when $\sin(2\beta)_{K_S\pi^0}$ is included~\cite{Fleischer:2008wb}. Relations involving isospin sum rules probe unambiguously the presence of non-SM physics, but the current reach of these tests is modest, owing to the large uncertainties on the $K^0\pi^0$ decay rates.  More than a hundred charmless $B$ decays studied by Belle and \babar\ continue providing useful information to constrain the hadronic unknowns~\cite{LeeTalk} and precise experimental information will be provided with large LHCb and super-flavor factories event samples. A similar situation is found for the ``polarization puzzles". Predictions of decay polarization amplitudes in $B\to VV$ decays support dominance of the longitudinal component. Belle and \babar\ measured polarization of about twenty decay modes and observed consistent and significant deviations in $b\to s$ penguin-dominated processes~\cite{VasseurTalk}. This was recently confirmed also in $B^0_s\to\phi\phi$ at CDF~\cite{Dorigo:2010by}. Several non-SM scenarios that could produce anomalous polarizations have been suggested. However, a more mundane explanation based on the presence of large 
 $1/m_b$ corrections is more probable~\cite{KaganTalk}. 

\section{Conclusions}

 The angle $\beta$ is determined with $5-10^\circ$ precision in many penguin-dominated channels, showing no significant discrepancy with the complementary (and more precise) results from charmonium decays. The angle $\alpha$ is extracted with somewhat lower precision and found consistent with values favored by the global CKM fits. Data are not yet sufficient for a solid extraction of $\gamma$ from penguin-dominated decays without some non-trivial theory assumptions. \par
Global agreement between experimental results and theory in charmless $B$ decays holds. However uncertainties in theory and data are still too large to allow exploration of  the full potential of these modes. This calls for significantly more precise results, as expected soon from the third generation flavor experiments. Increasing experimental precision will challenge the current theory limitations. Novel uses of theoretical tools able to cope with the problem of $1/m_b$ corrections would be highly desirable to fully exploit the experimental information that will be available.

\Acknowledgements
We thank Tim Gershon and Roger Forty for useful comments. This work was supported in part by the EU Marie Curie  IEF Grant. no. PIEF-GA-2009-252847 and by the Slovenian Research Agency.

\end{document}